\documentstyle[12pt,epsf]{article}
\title {Radiative Corrections to Moller Scattering  of
Polarized Particles}
\author {N. M. Shumeiko, J.G. Suarez
\and \it National Scientific and Education Center of Particle and
\and \it High Energy Physics attached to Byelorussian State University}
\def\beq{\begin{equation}}
\def\eeq#1{\label{#1}\end{equation}}
\def\barr#1{\begin{equation}\begin{array}{#1}\displaystyle}
\def\earr#1{\end{array}\label{#1}\end{equation}}
\def\bit{\begin{itemize}}
\def\eit{\end{itemize}}
\def\ben{\begin{enumerate}}
\def\een{\end{enumerate}}
\def\bce{\begin{center}}
\def\ece{\end{center}}
\def\bmi{\begin{minipage}}
\def\emi{\end{minipage}}
\def\btab{\begin{tabular}}
\def\etab{\end{tabular}}

\def\df{\stackrel{\rm def}{=}}
\def\lil#1{\hbox to\hsize{ #1 \hfil}}
\def\lilr#1#2{\hbox to\hsize{#1 \hfil #2 }}
\def\lir#1{\hbox to\hsize{\hfil #1 }}
\def\lic#1{\hbox to\hsize{\hfil #1 \hfil}}
\def\Z{\indent \indent }

\begin{document}
\maketitle
\begin {abstract}
Principal contributions to QED radiative effects for Moller
scattering of polarized particles are investigated both on the
Born level and taking into account radiative corrections (RC).
Scattering on the case of longitudinal and
transversal polarized targets is also considered.
 In a general way the expressions for differential cross section
  and polarization asymmetry (PA) have been defined,
 and  respective graphics for longitudinal and trans\-versal
 polarization asymmetries and also for radiative corrections
 to asymmetry are presented. All quantities
are presented in terms of covariant variables. The
ultrarelativistic approximation was applied for
calculations. The results of a computer run are presented.
\end {abstract}

\section{Introduction}
\Z For a wide serie of experiments in SLAC
(E-142,E-143,E-154,E-155), it is neccesary to know the
polarization
of electron beam. For this aim a single arm Moller polarimeter
is utilized. In order to extract (with desirable accuracy) beam
polari\-zation $p_b $ we need to calculate theoretically, taking
into account the experimental conditions, the quantities
$\sigma^{th}$ and $ A^{th}$ - theoretical meaning of cross
section
(CS) and polarization asymmetry. If the experimental data for
polarization asymmetry ($ A^{meas}$) and target polarization
($ p_t$) are known, the ratio
\begin{equation}
    A^{meas} = p_bp_tA^{th},
\end{equation}
 allows to find the polarization of electron beam.

\section{Method of Calculation}
\Z The Moller ($e^-+e^-\to e^-+e^-$) scattering CS
of order O($\alpha^{3}$) can be written in the form
(within the QED treatment)
\begin{equation}
  {\sigma} = {\sigma _ {o}} + {\sigma _{V}} + {\sigma _{R}},
\end{equation}
where each $ \sigma_ {o, V, R} \df d \sigma_ {o, V, R} / dy$,
and $y \df 1-E'/E $, where E(E') is the energy of the initial
(scattered) electron.
$\sigma_{o}$ is the Born (non radiative) contribution of order
O($\alpha^{2}$).
$\sigma_{V}$ is the contribution of the diagrams with an
additional virtual photon (V-contribution).
It is constitued by contributions of vacuum polarization
 (sum over all generations of leptons and quarks), vertex
renormalization and two-photon exchange diagrams.
$\sigma_{R}$ is the contribution of nonobservable
(internal bremsstrahlung) photon radiation (R-contribution).
$\sigma_{R}$ can be divided into three parts: $\sigma_{R}^{F}$
is finite when $k\to 0$ (k is the photon momentum),
$\sigma_{H}$ is the contribution of "hard photons", and it is also
finite, and $\sigma_{S}$ is the contribution of "soft photons".
The latter contains infrared divergences. All $\sigma_{R}^{F}$,
$\sigma_{H}$ and $\sigma_{S}$ may be calculated in a standard way
(see, for example, \cite{BS}-\cite{BaSh}). According to the
 method of $ref.\cite{BS}$, the
infrared divergence vanishes when the infrared divergent part of
$\sigma_{V}$ (denoted as $\delta_V^\lambda$ , see, for example,
\cite{BS}) is
summed with the infrared divergent part of $\sigma_{R}$
(denoted as $\delta_\lambda^R$, see, for example, \cite{BS}).

\Z As it is wellknown, the linear, exchangable and interfering
 diagrams give
contribution to Moller scattering. For this reason the
expression for differential cross section and polarization
asymmetry are presented as a sum of three parts:

\begin{equation}
  \frac{d\sigma}{dy}= \sum_{l}\frac{d\sigma_{l}}{dy},
\end{equation}
 and
\begin{equation}
   A= \sum_{l}A_{l},
\end{equation}
where $l$=$l$,$e$,$i$.
In our calculations the differential cross section is presented
as
\begin{equation}
\begin{array}{l}
\displaystyle
  \sigma = {\sigma _ {o} +
\frac{\alpha} {\pi}(\delta_{vert} + \delta_{vac})\sigma _ {o}} +
\sigma _ {amm}^u+ \sigma_{amm}^p+
\frac{\alpha}{\pi}(\delta_{k} + \delta_{2\gamma}^u)\sigma_{o}^u+
\\[0.5cm]
\displaystyle
\frac{\alpha} {\pi}(\delta_{k} +
\delta_{2\gamma}^p)\sigma_{o}^p + \sigma _{R}^{F,u} +
\sigma_{R}^{F,p} +
\frac{\alpha} {\pi}(\delta_{l} + \delta^{\lambda} +
\delta_{s}   +
\delta_{1}^{H})\sigma _ {o} + \sigma_{2}^{H,u} +
\sigma_{2}^{H,p},
\end{array}
\label{fff}
\end{equation}
where
$ \sigma_{o}=\sigma_{o}^u + \sigma_{o}^p$,
and $\sigma_{o}^{u,p}$ are the spin-independent
and spin-dependent Born contributions.
 $\delta_{vert}$ and $\delta_{vac}$ are the spin-independent
finite factorized contributions of vertex and vacuum polarization
corrections.
 $\sigma_{amm}^{u,p}$ are the spin-independent
  and spin-dependent contributions attributed to
   anomalous magnetic momentum.
$\delta_{2\gamma}^{u,p}$ are the
spin-independent and spin-dependent corrections due
 to two-photon exchange.
$\delta_{k}$ can be written in a standard way. For references
see \cite{Kah}.
$\sigma _{R}^{F,u,p}$  are the
spin-independent and spin-dependent cotributions of infrared free
part of R-contribution. $\delta_{l}$, $\delta_{s}$
$\delta_{1}^{H}$
and $\delta^{\lambda}$ are spin-independent factorized corrections
derived from the infrared divergent vertex contribution, finite
"soft-photon" contribution, finite "hard-photon" contribution
and the sum of infrared divergent parts in R and V contributions
($\delta^{\lambda}$).
$\sigma_{2,u}^{H}$ and $\sigma_{2,p}^{H}$ are the
spin-independent and spin-dependent corrections of infrared free
part of hard-photon contribution.
All contributions are defined in a standard way, (see, for
example, references \cite{BS,Kah}).

\Z The longitudinal and transversal polarization asymmetries are,
usually, defined as
\begin{equation}
A_{l}= {\sigma^{\uparrow \uparrow} - \sigma^{\uparrow \downarrow}
\over
\sigma^ {\uparrow \uparrow} + \sigma^ {\uparrow \downarrow}},
\end{equation}
and
\begin{equation}
A_{t}= {\sigma^{\uparrow \rightarrow} - \sigma^{\uparrow
\leftarrow}
\over \sigma^ {\uparrow \rightarrow} +
\sigma^ {\uparrow \leftarrow}},
\end{equation}
The arrows show the polarization of the beam and the target,
respectively.
The corrected polarization asymmetry is presented as
\begin{equation}
 A^{QED}= A_{o}(1+\frac{\alpha} {\pi}(\delta_{2\gamma}^{p}-
\delta_{2\gamma}^{u}))+\sigma_{amm}^{p}-
\sigma_{amm}^{u}+
\frac{{\sigma_{o}^{u}}{\sigma_{R}^{p}}-
{\sigma_{o}^{p}}{\sigma_{R}^{u}}}{(\sigma_{o}^{u})^{2}}+O(\alpha^{2}),
\end{equation}
where $A_{o}$ is the Born asymmetry, and
$A_{o} \df A_{lo}$ or $A_{o} \df A_{to}$.
$\sigma_{amm}^{p}$ and $\sigma_{amm}^{u}$
are the spin-dependent and spin-independent
contributions due to anomalous magnetic momentum.
$(\sigma_{o}^{u}\sigma_{R}^{p}-\sigma_{o}^{p}\sigma_{R}^{u})/
(\sigma_{o}^{u})^{2}$
is due to R-contribution.
Where $\sigma_{R}^{u}=\sigma_{R}^{F,u}+\sigma_{2,u}^{H}$ and
$\sigma_{R}^{p}=\sigma_{R}^{F,p}+\sigma_{2,p}^{H}$.

\Z The radiative correction to asymmetry is defined as
\begin{equation}
     \delta^{QED}= \frac {A^{QED}} {A_{o}} - 1.
\end{equation}
\section{Results}
\Z In this section the results of calculations
   of polarization asymmetry both for longitudinally
  and transversally polarized target and radiative
  corrections to asy\-mme\-try are presented.
 Numerical calculations have been carried out
 with the help of FORTRAN code created by the authors.
  For analytical
 calculations REDUCE code has been utilized. Graphics
 for asymmetries and radiative corrections can be seen in
 figures 1-4. Obtained results are presented in table 1.
\begin{table}
\begin{center}
\begin{tabular}{||c|cccc|cc||}
\hline
$y$&$A_{ol}$&$A_{ot}$&$A_{l}$&$A_{t}$&$\delta_{l} (\%)$&$\delta_{t} (\%)$\\

\cline{1-7}
 .10&-.41282&.13334E-02&-.48454&.11093E-02&17.374&-16.807\\
 .20&-.66387&.29711E-02&-.72840&.27305E-02&09.720&-08.943\\
 .30&-.81058&.43023E-02&-.86688&.40466E-02&06.946&-05.147\\
 .40&-.88520&.51717E-02&-.93473&.49056E-02&05.595&-05.194\\
 .50&-.90438&.55191E-02&-.95268&.52324E-02&05.340&-05.823\\
 .60&-.87293&.53009E-02&-.92673&.49923E-02&06.224&-05.233\\
 .70&-.78220&.44700E-02&-.84584&.41523E-02&08.136&-07.233\\
 .80&-.61358&.30365E-02&-.68504&.26956E-02&11.647&-11.224\\
 .90&-.32977&.11483E-02&-.38357&.79126E-02&16.315&-31.092\\
 .99&-.04891&.63291E-02&-.05877&.15068E-02&20.148&-338.07\\
\hline
\end{tabular}
\end{center}
\caption{
 Results of calculations of Born asymmetries
 $A_{ol,ot}$, corrected asymmetries $A_{l,t}$
 and radiative corrections to asymmetry $\delta_{l,t}$ for both
 longitudinally and transvesally polarized tagets
 correspondingly, for SLAC kinematics. E=50 Gev.}
\label{tab4}
\end{table}

\section{Conclusions}
\Z The calculation of the differential cross section and
polarization asymmetry to Moller scattering
will in the future allow to find beam polarization. For this aim
it is neccesary to measure polarization asymmetry with high
accuracy. Iteration procedure for theoretical
calculation of polarization asymmetry can be improved,
and that will lead to the increase of the accuracy when beam
polarization is measured.
As we can see, the corrections without cuts are very
large (in any ocasion about 20\%), although they
will be considerably reduced when the experimental
cuts will be taken into account.
Analysis of the corrected polarization asymmetries in the case of
longitudinally and transversally polarized targets show that in a
wide kinematical
range corrections are considerable.
In this paper  the results of calculations of differencial cross section
 and polarization asymmetry have been presented
taking into account only the contribution of linear diagrams.
Results, in which  the contribution of exchangable
and interfering diagrams are considered
will be presented in the future.
For a more accurate calculation of polarization asymmetry
it is neccesary to utilize a Monte Carlo generator for Moller
scattering, and  the contribution
of correction due to external radiation of beam particles
within the target must be taken into account. This is an object
for future investigations.

\begin{figure}[p]
\vspace{0.5cm}
\hspace{0.5cm}
\unitlength 1mm
\begin{picture}(40,40)
\put(15,-15){
\epsfxsize=12cm
\epsfysize=12cm
\epsfbox{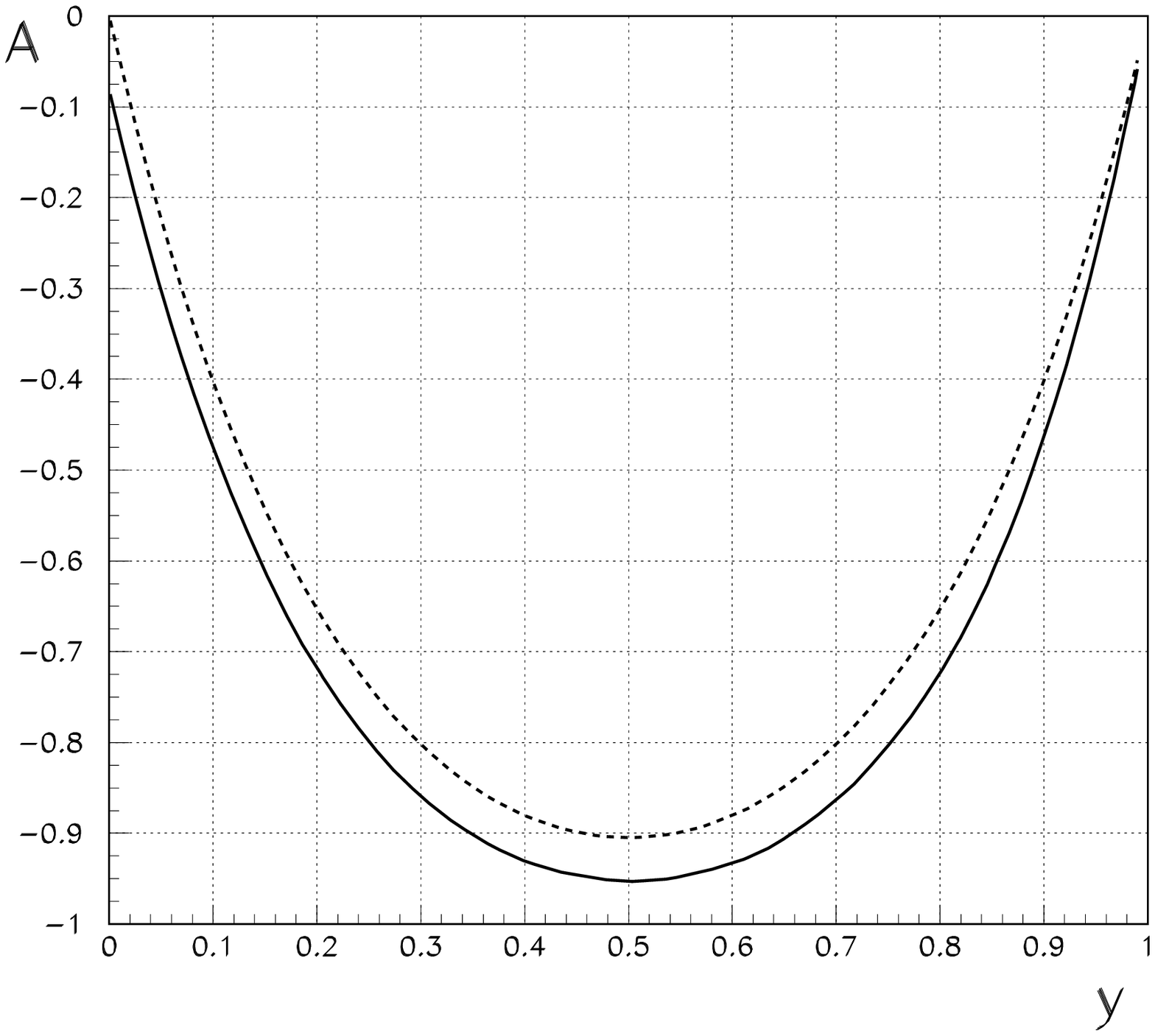}
}
\end{picture}
\caption{\protect\it y-dependence of the born asymmetry (dashed line)
and corrected asymmetry (solid line) in the polarization Moller
scattering for SLAC kinematics. E=50Gev.Longitudinally
polarized target.}
\end{figure}

\begin{figure}[p]
\vspace{4.2cm}
\hspace{0.5cm}
\unitlength 1mm
\begin{picture}(40,40)
\put(15,-20){
\epsfxsize=12cm
\epsfysize=12cm
\epsfbox{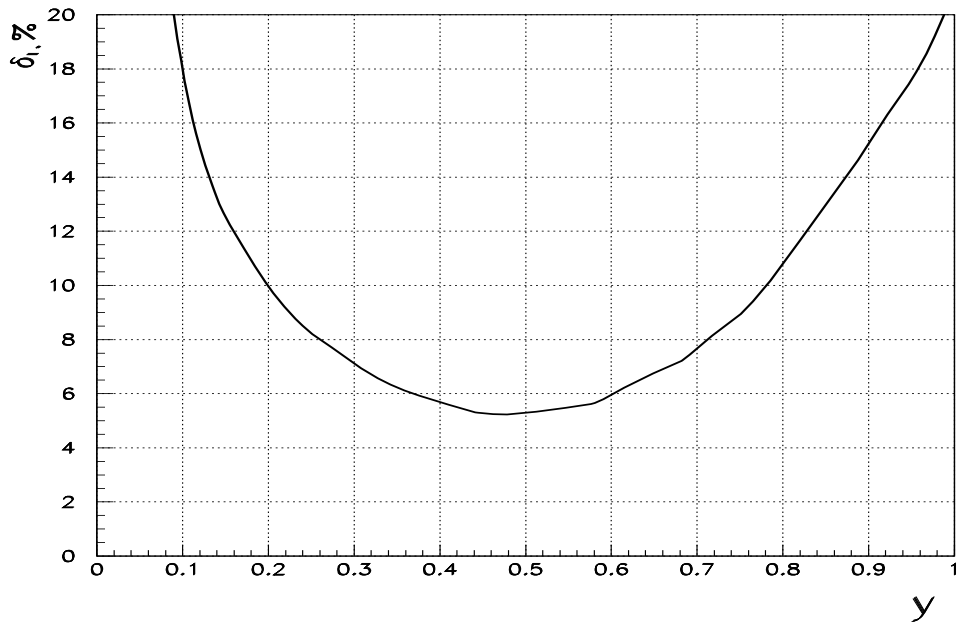}
}
\end{picture}
\caption{\protect\it The QED radiative corrections to asymmetry
without experimental cuts for longitudinally polarized target.}
\end{figure}

\begin{figure}[p]
\vspace{0.5cm}
\hspace{0.5cm}
\unitlength 1mm
\begin{picture}(40,40)
\put(15,-15){
\epsfxsize=12cm
\epsfysize=12cm
\epsfbox{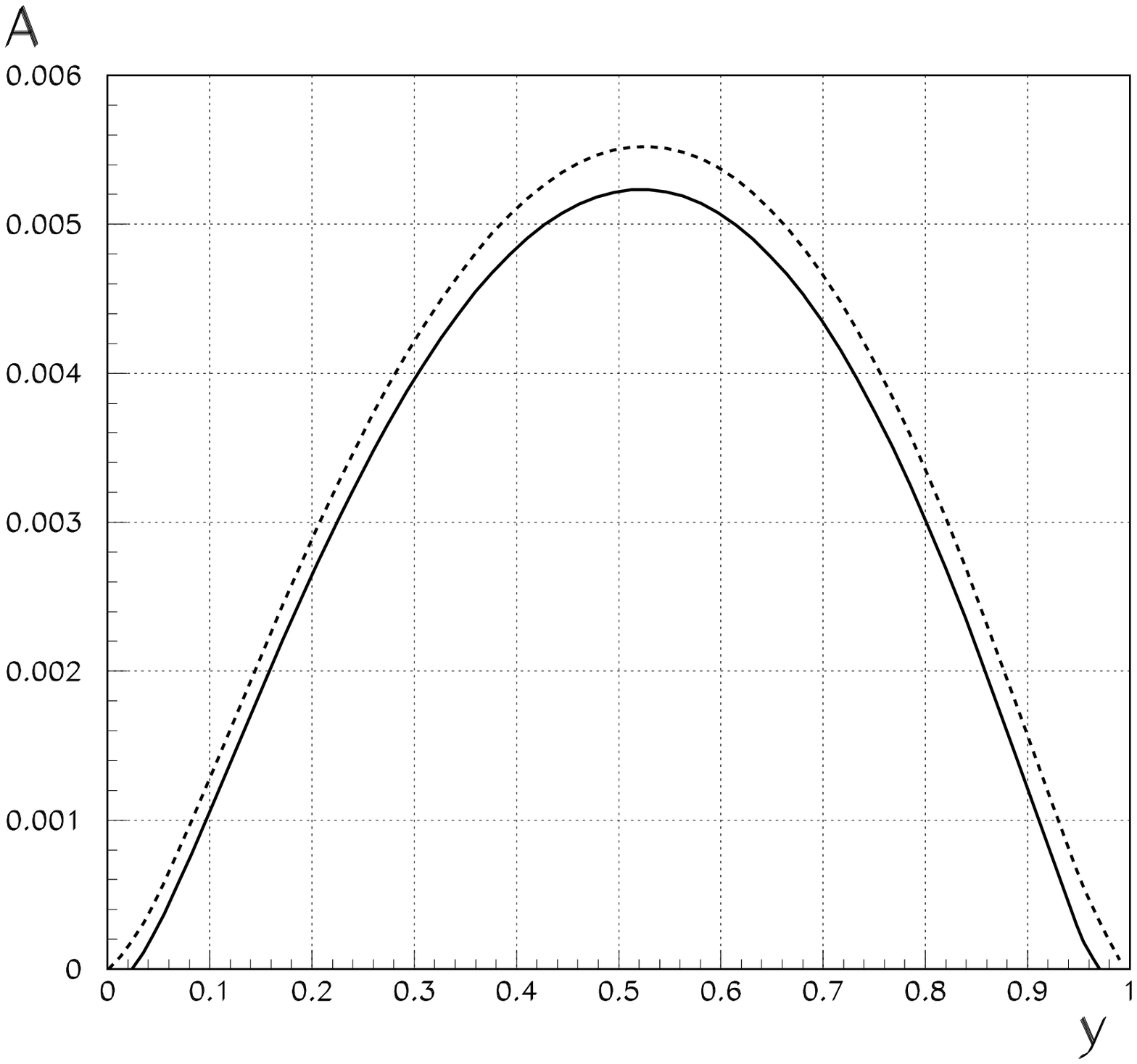}
}
\end{picture}
\caption{\protect\it y-dependence of the born asymmetry (dashed line)
and corrected asymmetry (solod line) in the polarization Moller
scattering for SLAC kinematics. E=50Gev.transversally
polarized target.}
\end{figure}

\begin{figure}[p]
\vspace{4.2cm}
\hspace{0.5cm}
\unitlength 1mm
\begin{picture}(40,40)
\put(15,-20){
\epsfxsize=12cm
\epsfysize=12cm
\epsfbox{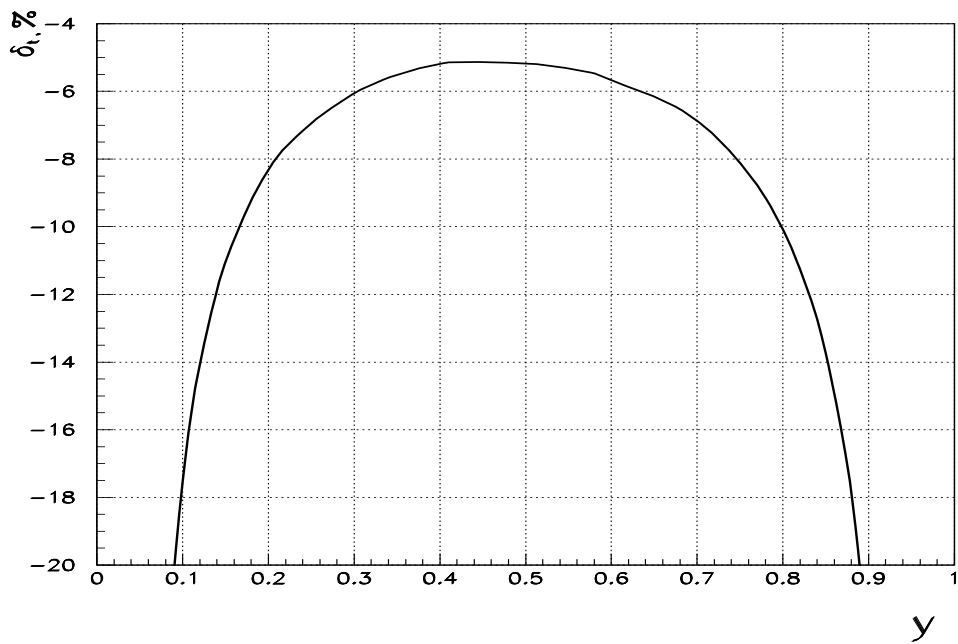}
}
\end{picture}
\caption{\protect\it The QED radiative corrections to asymmetry
without experimental cuts for transversally polarized target.}
\end{figure}
\label{Fg2}
\end{document}